\documentclass[prd,aps,twocolumn,nofootninbib]{revtex4}

\newcommand{\parfrac}[2]{\left(\frac{#1}{#2}\right)} %

\newcommand{\kev}{\ensuremath{\;\mathrm{keV}}} %
\newcommand{\keV}{\ensuremath{\;\mathrm{keV}}} %
\newcommand{\dm}{{\textsc{dm}}} 
\newcommand{\numsm}{$\nu$MSM\xspace} 
\newcommand{\xmm}{\textit{XMM-Newton}} 
\newcommand{\chan}{\textit{Chandra}} 
\newcommand{\suza}{\textit{Suzaku}} 

\newcommand{\fov}{\mathrm{fov}} 
\newcommand{\eff}{\mathrm{eff}} 

\usepackage{units} 

\usepackage{tabularx}

\usepackage{booktabs} 
\newcommand {\otoprule }{\midrule [\heavyrulewidth]} 

 \usepackage{graphicx,grffile}

\usepackage{paralist,color,xspace,eurosym,amsmath,amssymb,amsfonts,bm}
\usepackage{array,hhline}

\usepackage[colorlinks=false,bookmarks=false]{hyperref}

\begin{document}
\title{Potential of LOFT telescope for the search of dark matter}

\author{A.~Neronov$^{1}$, A.~Boyarsky$^{2,3,4}$, D.~Iakubovskyi$^{4,5}$ and O.~Ruchayskiy$^{6,3}$\\
  $^1${\small ISDC Data Centre for Astrophysics, Department of Astronomy,}\\
  {\small University of Geneva, Ch. d'Ecogia 16, 1290, Versoix, Switzerland}\\
  $^2${\small Instituut-Lorentz for Theoretical Physics, Universiteit Leiden,}\\
  {\small Niels Bohrweg 2, Leiden, The Netherlands}\\
  $^3${\small Ecole Polytechnique F\'ed\'erale de Lausanne,}\\
  {\small FSB/ITP/LPPC, BSP
    720, CH-1015, Lausanne, Switzerland}\\
  $^4${\small Bogolyubov Institute of Theoretical Physics,}\\
  {\small Metrologichna Str. 14-b, 03680, Kyiv , Ukraine}\\
  $^5${\small National University ``Kyiv-Mohyla Academy'',}\\
  {\small Skovorody Str. 2, 04070, Kyiv, Ukraine}\\
  $^6${\small CERN Physics Department, Theory Division,}\\
  {\small CH-1211 Geneva 23, Switzerland} %
}\date{}

\begin{abstract}
  \emph{Large Observatory For X-ray Timing} (LOFT) is a next generation X-ray
  telescope selected by European Space Agency as one of the space mission
  concepts within the ``Cosmic Vision'' programme. The  Large Area Detector on board of LOFT will be a collimator-type telescope with an
  unprecedentedly large collecting area of about $\unit[10^5]{cm^2}$ in the
  energy band between 2 and 100~keV. We demonstrate that LOFT
  will be a powerful dark matter detector, suitable for the search of
  the X-ray line emission expected from decays of light dark matter particles
  in galactic halos.  We show that LOFT will have sensitivity for dark matter line search
  more than an order of magnitude higher than that of all existing X-ray
  telescopes. In this way, LOFT will be able to provide a new insight into the fundamental problem of the nature of dark matter.
\end{abstract}

\maketitle

\section{Introduction}

The nature of dark matter (DM) is one of the most intriguing questions of
modern physics. Mass content of galaxies and galaxy clusters, growth of
density fluctuations through the cosmic history, large scale structure of the
Universe -- all point towards the existence of new substance, the DM, which constitutes some 80\% of the total mass content
of the Universe~\cite{Planck-cosmo}.  If DM is made of particles,
these particles are not among the known ones.  Phenomenologically little is
known about properties of DM particles:
\begin{itemize}[--]
\item Their overall density is $\Omega_\text{DM}h^2 = 0.1196 \pm
  0.0031$~\cite{Planck-cosmo};
\item The mass of any fermionic DM is limited from below by the
  ``Tremaine-Gunn bound''~\cite{Tremaine:79}, while for bosons such a limit is
  significantly lower~\cite{Madsen:90,Madsen:91}.
\item Dark matter particles are not necessarily stable, but their lifetime
  should significantly exceed the age of the Universe (see
  e.g.~\cite{Boyarsky:08b,Bertone:07,Essig:2013goa});
\item DM particles should have become non-relativistic sufficiently early in
  the radiation-dominated epoch (although a sub-dominant fraction might have
  remained relativistic much later~\cite{Boyarsky:08c}).
\end{itemize}
Depending on the nature of interactionof DM particles  with ordinary matter today, the DM  can
have different astrophysical signatures~(see e.g.\
\cite{Cirelli:2012tf,Profumo:2013yn}). Two
main classes of DM particle candidates are considered:  \emph{annihilating} and \emph{decaying}.

A lot of attention has been devoted to a class of annihilating DM candidates called weakly interacting massive particles (WIMPs)
(see e.g.~\cite{Bertone:04,Feng:10} for review).  These hypothetical particles
are assumed to interact with ordinary matter with roughly electroweak strength
and have masses in $\mathcal{O}(1{-}10^3)$~GeV to provide the correct DM
abundance.  Due to their large mass and interaction strength these particles
should be stable and astrophysical signature of their annihilation products is
an important scientific goal of many cosmic
missions~\cite{Strigari:2012gn,Profumo:2013yn}. In particular, $\gamma$-rays
from DM annihilation are extensively searched with $\gamma$-ray telescopes \cite{hess,fermi}.

There is a large class of DM candidates that interact with the ordinary
particles super-weakly (i.e.\ significantly weaker than neutrinos).  These
include: extensions of the SM by right-handed neutrinos
\cite{Dodelson:93,Asaka:05a,Lattanzi:07}, models with extra dimensions and
string-motivated models~\cite{Conlon:07},
gravitinos~\cite{Takayama:00,Buchmuller:07},
axions~\cite{Sikivie:2006ni,Kawasaki:2013ae},
axinos~\cite{Covi:01,Choi:2013lwa} (see
e.g.~\cite{Taoso:07,Steffen:2008qp,Feng:10} for reviews).  These candidates
are as possible as WIMPs and from many points of view are very compelling.
The feeble interaction strength of these DM candidates means that unlike
WIMPs: (i) their mass is not restricted to the GeV region; and (ii) they can
\emph{decay} into the SM particles.  The fermionic DM candidates (such as
sterile neutrino, gravitino, axino) posses a 2-body radiative decay channel:
${DM} \to \gamma + \nu$, while bosonic DM candidates (such as e.g.\ axion or
Majoron) can decay into two photons. These 2-body decays produce photons with
energy $E_\gamma = \frac 12 M_{DM}c^2$. The cosmologically long lifetime makes
the intrinsic width of such a line negligible.  This provides a clear
  observational signature of decaying DM candidates: a narrow spectral line
in spectra of DM-dominated objects, correlated with DM density
distribution.

Search of the DM decay signal in the keV--MeV mass range was conducted
using \xmm~\cite{Boyarsky:05,Boyarsky:06b,Boyarsky:06c,Watson:06,Boyarsky:06d,Boyarsky:07a,Loewenstein:12},
\textit{Chandra}~\cite{Riemer-Sorensen:06a,Riemer-Sorensen:06b,Boyarsky:06e,Riemer-Sorensen:09,Loewenstein:09,Boyarsky:10b},
\textit{Suzaku}~\cite{Loewenstein:08,Kusenko:12}, \textit{Swift}~\cite{Mirabal:10b},
\textit{INTEGRAL}~\cite{Yuksel:07,Boyarsky:07b} and HEAO-1~\cite{Boyarsky:05} cosmic missions, as well as rocket-borne X-ray
microcalorimeter~\cite{Boyarsky:06f}. 
Observations
of extragalactic diffuse X-ray background~\cite{Boyarsky:05,Abazajian:06b};
galaxy clusters~\cite{Boyarsky:06b,Riemer-Sorensen:06b,Boyarsky:06e}; Milky Way, Andromeda (M31) and Triangle (M33)
galaxies~\cite{Abazajian:06b,Watson:06,Boyarsky:06b,Boyarsky:06c,Boyarsky:06d,Boyarsky:07b,Yuksel:07};
dwarf spheroidal satellites of the Milky Way
\cite{Boyarsky:06c,Boyarsky:06f,Boyarsky:10b,Riemer-Sorensen:09,Loewenstein:08,Loewenstein:12,Kusenko:12}
allowed to put important constraints on particle physics parameters,
establishing lower bounds on decaying DM lifetime to be at least 8 orders of
magnitude longer than the age of the Universe~\cite{Boyarsky:08b} (see
also~\cite{Essig:2013goa} for extension for higher energies).
 Table~\ref{tab:bounds-summary} summarizes existing works that put bounds on
decaying DM from observations of individual objects. In this Table, we do not mention 
  the claim~\cite{Prokhorov:10} that the intensity of the Fe~XXVI
  Lyman-$\gamma$ line at $8.7\kev$, observed in~\cite{Koyama:06} cannot be
  explained by standard ionization and recombination processes, and that the
  dark matter decay may be a possible explanation of this apparent excess. Spectral
  resolution of current missions does not allow to reach any
  conclusion. However, barring an \emph{exact} coincidence between energy of
  decay photon and Fe~XXVI Lyman-$\gamma$, this claim may be tested with the
  new missions, discussed in e.g.~\cite{Boyarsky:12c}.

\begin{table*}
    \begin{tabularx}{\textwidth}{l|l|l|c}
      \toprule
      Ref. & Object & Instrument & Cleaned exp, ks  \\
      \hline
      \otoprule
      \cite{Boyarsky:05} & Diffuse X-ray background & HEAO-1, \xmm & 224, 1450  \\
      \midrule
      \cite{Boyarsky:06b} & Coma \& Virgo galaxy clusters & \xmm & 20, 40  \\
      \midrule
      \cite{Boyarsky:06c} & Large Magellanic Cloud & \xmm & 20  \\
      \midrule
      \cite{Riemer-Sorensen:06a} & Milky Way halo & \chan/ACIS-S3 & Not specified  \\
      \midrule
      \cite{Watson:06} & M31 (central $5'$) & \xmm & 35  \\
      \midrule
      \cite{Riemer-Sorensen:06b} & Abell~520 galaxy cluster & \chan/ACIS-S3 & 67  \\
      \midrule
      \cite{Boyarsky:06d} & Milky Way halo, Ursa Minor dSph & \xmm & 547, 7  \\
      \midrule
      \cite{Abazajian:06b} & Milky Way halo & \chan/ACIS & 1500  \\
      \midrule
      \cite{Boyarsky:06e} & Galaxy cluster 1E~0657-56 (``Bullet'') & \chan/ACIS-I & 450  \\
      \midrule
      \cite{Boyarsky:06f} & Milky Way halo & X-ray microcalorimeter & 0.1  \\
      \midrule
      \cite{Yuksel:07} & Milky Way halo & INTEGRAL/SPI & 5500  \\
      \midrule
      \cite{Boyarsky:07a} & M31 (central $5-13'$) & \xmm/EPIC & 130  \\
      \midrule
      \cite{Boyarsky:07b} & Milky Way halo & INTEGRAL/SPI & 12200 \\
      \midrule
      \cite{Loewenstein:08} & Ursa Minor & \suza/XIS & 70  \\
      \midrule
      \cite{Riemer-Sorensen:09} & Draco dSph & \chan/ACIS-S & 32  \\
      \midrule
      \cite{Loewenstein:09} & Willman~1 & \chan/ACIS-I & 100  \\
      \midrule
      \cite{Boyarsky:10b} & M31, Fornax, Sculptor & \xmm/EPIC , \chan/ACIS & 400, 50, 162   \\
      \midrule
      \cite{Mirabal:10a} & Willman~1 & \chan/ACIS-I & 100\\
      \midrule
      \cite{Mirabal:10b} & Segue~1 & \textit{Swift}/XRT & 5  \\
      \midrule
      \cite{Borriello:11} & M33 & \xmm/EPIC & 20-30  \\
      \midrule
      \cite{Watson:11} & M31 ($12-28'$ off-center)& \chan/ACIS-I & 53 \\
      \midrule
      \cite{Loewenstein:12} & Willman~1 & \xmm & 60 \\
      \midrule
      \cite{Kusenko:12} & Ursa Minor, Draco & \suza/XIS & 200, 200 \\
      \bottomrule
      \hline
    \end{tabularx}
\caption{Summary of existing X-ray observations of different objects performed
  by different groups.
}
  \label{tab:bounds-summary} 
  \end{table*}

In what follows we argue that a next-generation X-ray mission  Large Observatory For x-ray Timing (LOFT) will provide a crucial improvement in the sensitivity for the search of decaying DM in X-rays.  LOFT mission is under a study at the European Space Agency (ESA) as one of the five medium mission candidates for the launch after 2020 in the framework of the ``Cosmic Vision'' program of ESA. Further details on the LOFT mission could be found at \url{http://www.isdc.unige.ch/loft/}. We also show that LOFT will  have a capability to explore almost the entire parameter space of one of the most often discussed models of decaying DM, the neutrino minimal extension of the Standard Model of particle physics ($\nu$MSM) \cite{Asaka:05a} if converted into a dedicated DM detection experiment (e.g. toward the end of the mission) aimed at ultra-deep exposure of the most favourable (massive, relatively compact) nearby DM halo.

\section{Strategy of searching for decaying dark matter}
\label{sec:searching-decaying-dark-matter}

The number of photons from DM decay is proportional to the DM
  column density $S_{DM} = \int\rho_{DM}(r)dr$ (integrated DM distribution along
the line of sight) and not to the $\int\rho^2_{DM}(r)dr$ (as in the case of
the annihilating DM).  As it turns out, this signal is very weakly dependent
on the virial mass of the DM halo and on the assumed dark matter density
profile~\cite{Boyarsky:06c,Boyarsky:09b,Boyarsky:09c}.  Moreover, for objects
that cover the whole field of view of the telescope, the expected DM decay
flux is independent on the distance to the object. As a result a vast
variety of DM-dominated objects (nearby galaxies and galaxy clusters) produce a
comparable decay signal.  Therefore
\begin{inparaenum}[\it (i)]\item one has a freedom of choosing the
  observational targets, avoiding complicated astrophysical backgrounds;
\item if a candidate line is found, its surface brightness profile may be
  measured, distinguished from known atomic  lines  and compared among several objects with the same expected signal
  (see e.g.~\cite{Boyarsky:10b}).  This allows to distinguish the decaying DM
  line from astrophysical backgrounds.
\end{inparaenum}
The case of the astrophysical search for decaying DM has been presented in the
recent White Papers~\cite{Abazajian:09a,denHerder:09}.

With intrinsic width of the decay line being negligible, its broadening is
determined entirely by the virial velocity of DM particles, confined to in a
halo: {$E/\Delta E \simeq c/v_{\rm vir}$}. This number ranges from $10^2$ for
galaxy clusters to $10^4$ for dwarf spheroidal galaxies. The spectral
resolution of modern X-ray instruments is insufficient to resolve this line
(with an exception of INTEGRAL's spectrometer SPI,
see~\cite{Boyarsky:07b}). The narrow line is detected on top of a continuum
background. This background has two main contributions -- astrophysical and
instrumental. The astrophysical background is a continuum thermal and
non-thermal emission form the source medium: interstellar/intraclusted medium
of galaxies and galaxy clusters, and from the set of isolated sources, like
X-ray binaries situated in the source host galaxy or galaxy cluster plus the
Cosmic X-ray background (CXB) \cite{Churazov:07} within the instrument's
Field-of-View (FoV). The instrumental background is produced by the charged
particles passing through the detector and by the electronic noise. The line
signal is centered on the reference line energy $E$ and is smeared over the
energy range $\sim (2-3)\times \Delta E$ where $\Delta E$ is a spectral
resolution. The amount of background accumulated in this energy bin is
proportional to the bin width $\Delta E$. Thus, improvement of the energy
resolution results in the decrease of the background and, as a consequence,
improvement of the sensitivity of the instrument for the line detection.

\begin{figure}
\includegraphics[width=\linewidth]{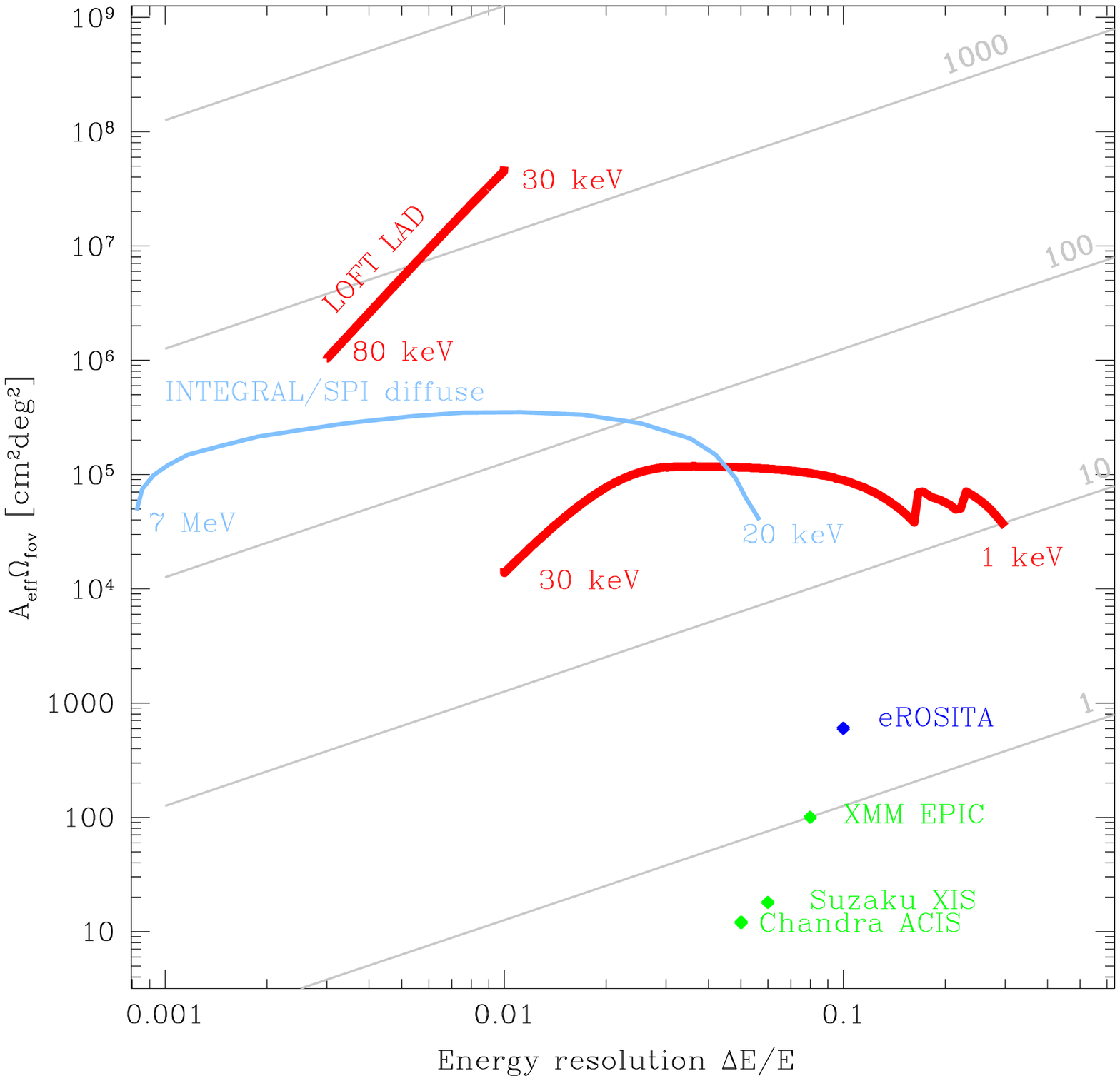}
\caption{Sensitivity of
  X-ray telescopes for the dark matter decay line detection in terms of the
  ``energy resolution vs. grasp'' diagram (c.f.~\cite{Boyarsky:06f}). Two red
  solid curves correspond to the LAD detector in two different observation
  modes: observations of localized sources of the angular extent i~$\gtrsim
  1^\circ$ range and observations of the large angular scale diffuse emission
  from the Milky Way with the steradian-sized FoV of LAD at higher
  energies. Dashed line shows the grasp of the WFM detector of LOFT.  Inclined
  grey lines with marks in 1-100 range show improvement of the sensitivity for
  the line search due to the increase of effective area / FoV and improvement
  of energy resolution. Level ``1'' corresponds to average parameters of the
  \xmm\ EPIC camera. Notice that points on the curves for LOFT and
  INTEGRAL/SPI correspond to different energies, from 1 to 100~keV and from
  20~keV to 7~MeV, respectively.}
\label{fig:diagram}
\end{figure}

The significance of the line signal from a diffuse source increases with the
collection area of the detector. It is proportional to the product of the
effective area, $A_\eff$ on the solid angle subtended by the FoV (for those DM
halos that have angular size larger than the FoV) that is to the ``grasp''
$A_\eff\Omega_\fov$ of the instrument \cite{Boyarsky:06f}.
Comparison of potential of different instruments for the detection of DM decay
line could be conveniently presented in terms of ``energy resolution vs.\
grasp'' diagram \cite{Boyarsky:06f}, as shown in Fig.~\ref{fig:diagram}.

In this figure, the inclined lines show the ``equal sensitivity'' sets of
instrumental characteristics. Indeed, the signal-to-noise ratio for the DM
decay line sensitivity improves as $R\propto \sqrt{A_{\rm eft}\Omega_{\rm
    fov}/\Delta E}$, so that the lines ``grasp''$\propto$``energy resolution''
correspond to instruments which provide the same signal-to-noise ratio if they
operate in the same energy band.  One could define $R$ as a ``figure of
merit'' for the {weak} line search, {see
  e.g.~\cite{Kaastra-ixo-spectroscopy}}.  We have arbitrarily fixed $R=1$ for
the parameter choice corresponding to the averaged over the energy band
characteristics of the EPIC camera of the \xmm\ telescope~\cite{Boyarsky:06f}.

The comparison shown in Fig. \ref{fig:diagram} adopts an assumption that the level of background in different instruments is comparable. This is true if the background on top of which the DM signal is searched is the CXB. However, if the background is of instrumental nature, the comparison of different instruments has to include an additional parameter, which is the level of background. We include this parameter in our considerations below.

\begin{table}
\begin{tabular}{lll}
  \hline
  Parameter & Requirement & Goal \\
  \hline
  Energy range & 2--30~keV  & 1--40 keV  \\
  & 2--80~keV \cite{Feroci:12,LOFT_LAD_SPIE:12}& 1--80 keV  \cite{LOFT_LAD_SPIE:12} \\
  \vspace{-0.2cm}\\
  Eff. area & 12.0~m$^2$ (2--10 keV) & 15~m$^2$ (2--10 keV) \\
  & 1.3~m$^2$ (@30 keV) & 2.5~m$^2$ (@30 keV) \\
  \vspace{-0.2cm}\\
  $\Delta$E & $<$260 eV & $<$180 eV \\
  (FWHM, @6 keV)  &  &  \\
  \vspace{-0.2cm}\\
  FoV (FWHM) & $<$60 arcmin & $<$30 arcmin \\
  \hline
\end{tabular}
\caption{Scientific requirements for the LOFT LAD instrument 
  (from~\cite{Feroci:12,LOFT_LAD_SPIE:12}).
  The energy range
  of LOFT LAD detector can be extended beyond 30~keV (the \emph{nominal}
  range) to the energies up to 80~keV (see~\cite{LOFT_LAD_SPIE:12} for the
  latter number). At those higher energies the LAD collimator becomes more and
  more transparent to X-rays~\cite{Feroci:12}.}
\label{t:scireq}
\end{table}

\section{LOFT characteristics relevant for DM detection}
\label{sec:loft}

\begin{figure}
\includegraphics[width=\linewidth]{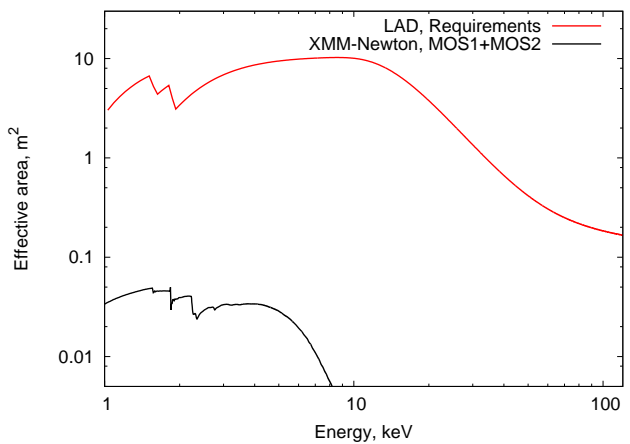}
  \caption{Characteristics of the LAD Effective areas for LAD instrument in
    the ``Requirements'' payload. For comparison the effective area of
    combined EPIC MOS1 + MOS2 cameras of \xmm\ is shown in black.}
\label{fig:LAD_area_Requirements_all_vs_single}
\end{figure}

The main instrument on board of LOFT will be the Large Area Detector (LAD). LAD will be an
X-ray telescope with effective collection area $A_{\rm eft}\simeq
10$~m$^2$ (see Fig.~\ref{fig:LAD_area_Requirements_all_vs_single}) sensitive
in the 2-80~keV energy range \cite{Feroci:12}. LAD will be composed of the Silicon Drift Detectors (SDD) with energy
resolution below 300~eV. The SDDs will be covered by microchannel plate
collimators providing the Field of View of $1^\circ$ in the energy range below
$\simeq 30$~keV and becoming increasingly transparent to X-rays at higher
energies up to 80~keV \cite{Feroci:12,LOFT_LAD_SPIE:12}. 

\begin{figure}
\includegraphics[width=\linewidth]{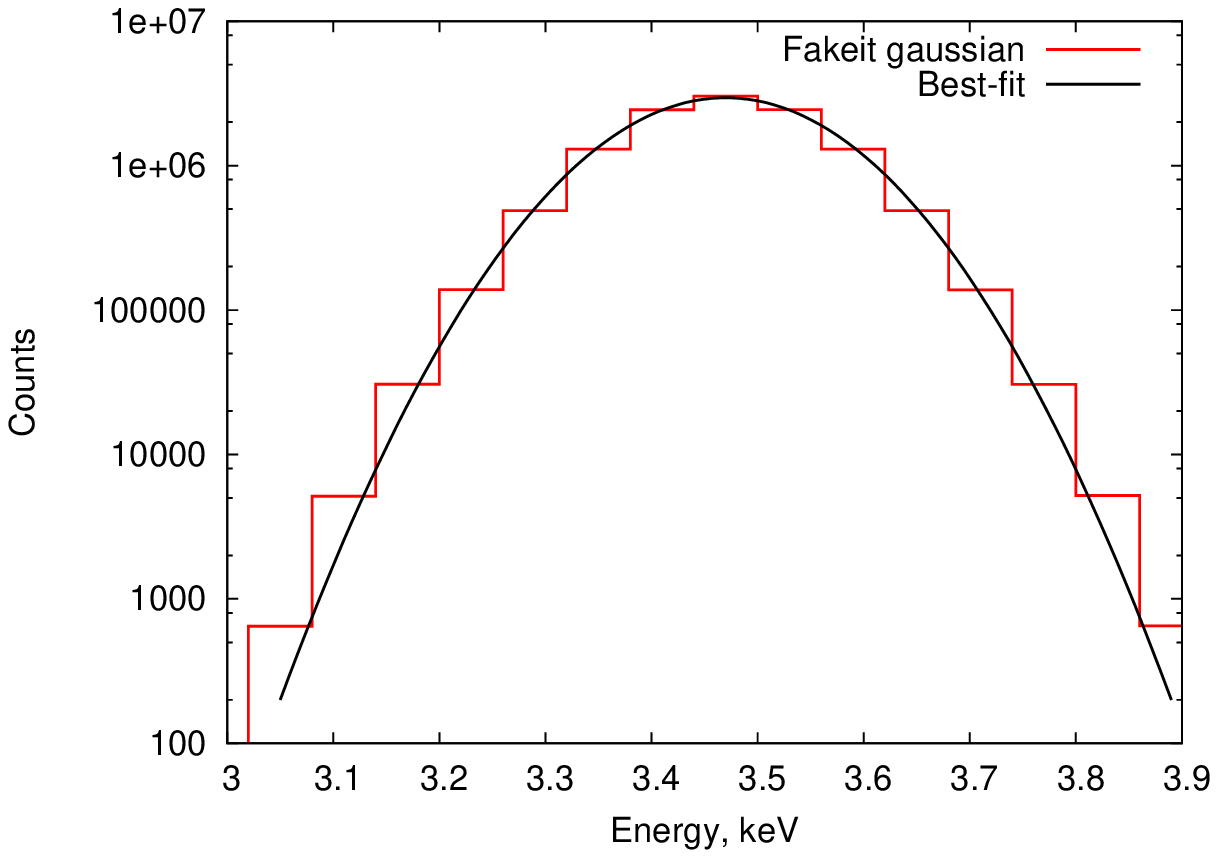}
\includegraphics[width=\linewidth]{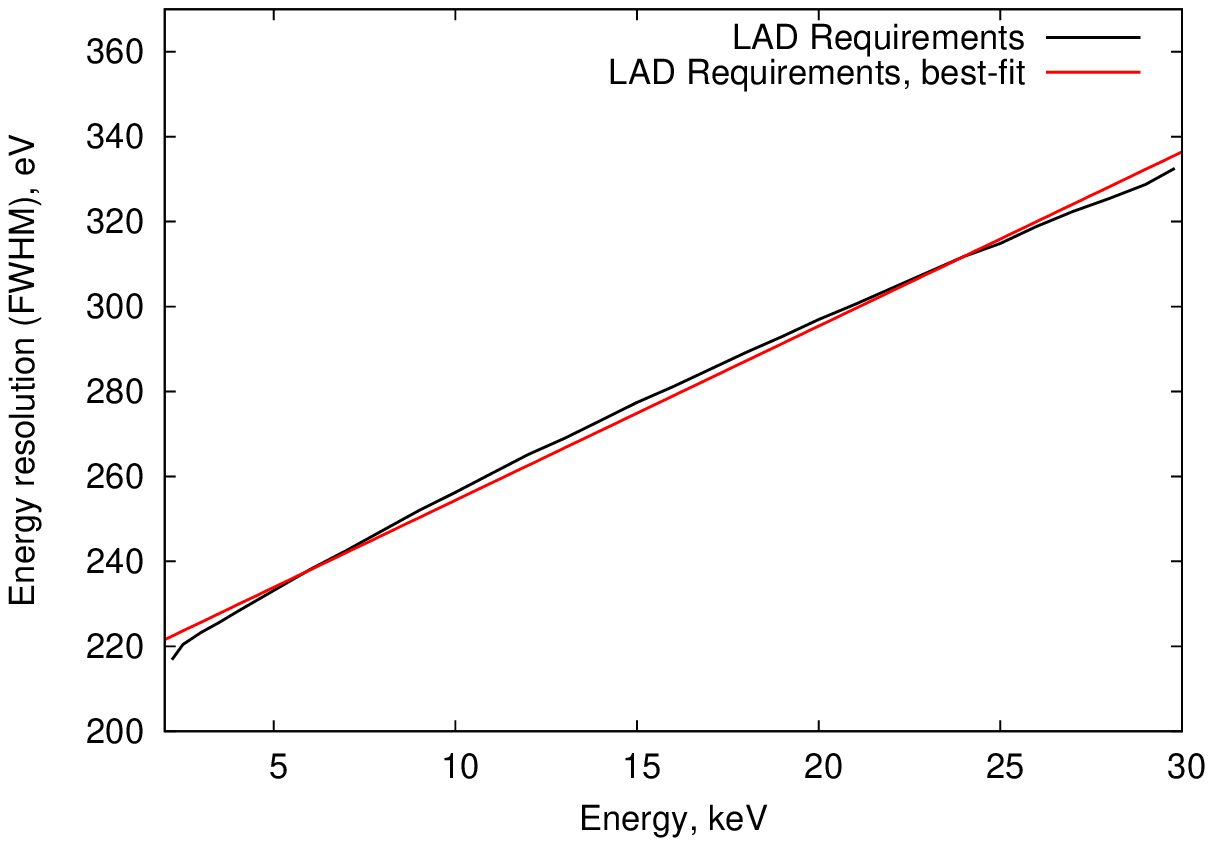}
\caption{\emph{Left}: an example of simulated line at 3.5 keV, together with
  its best-fit \texttt{gaussian} model used to calculate FWHM.  \emph{Right:}
  LAD energy resolution for ``Requirements'' payload as function of energy and its best fit (\ref{eq:LAD-energy-resolution-best-fit})
  calculated from our simulations. 
}
\label{fig:LAD_energy_resolution}
\end{figure}

The energy resolution, of LAD is determined by the characteristics of the silicon detectors and of the detector electronics
\cite{Feroci:12}.  Using the response functions of the LOFT
satellite\footnote{Available at~\cite{LAD-responses-background}.}, we
simulated narrow line at different energies and then approximated the obtained
spectrum by the Gaussian profile (see left panel of the
Fig.~\ref{fig:LAD_energy_resolution}). The obtained best-fit value of Gaussian
dispersion is then used to calculate FWHM. The results are shown in the right
in Fig.~\ref{fig:LAD_energy_resolution}. They can be approximated as a linear
function of energy: 
\begin{equation}
  \text{FWHM}(E) = 0.213~\keV + 4.10 \times
  10^{-3} \frac{E}{\unit{keV}}.\label{eq:LAD-energy-resolution-best-fit}
  \end{equation}

Our analysis considers two possible LOFT configurations\footnote{http://www.isdc.unige.ch/loft}: ``Requirements''
and ``Goal''. Parameters of each configuration are summarised in Table \ref{t:scireq}.

\begin{figure}[!t]
{\includegraphics[height=\linewidth,angle=-90]{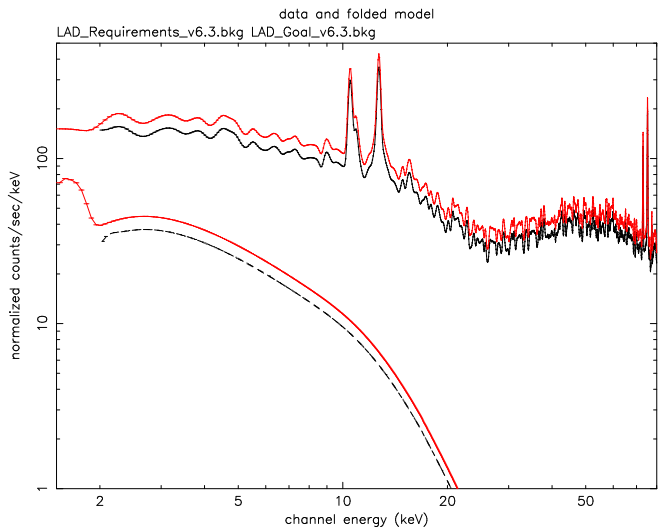}}
\caption{ Background of LAD instrument
  (compared to the CXB, lower curve). The instrumental component has been obtained using
  \texttt{LAD\_Requirements\_v6.3.bkg} background file from ISDC LAD response
  and background page~\protect\cite{LAD-responses-background}.}
\label{fig:LOFT-internal-vs-extragal-10ks}
\end{figure}

\begin{figure}
 {\includegraphics[width=\linewidth]{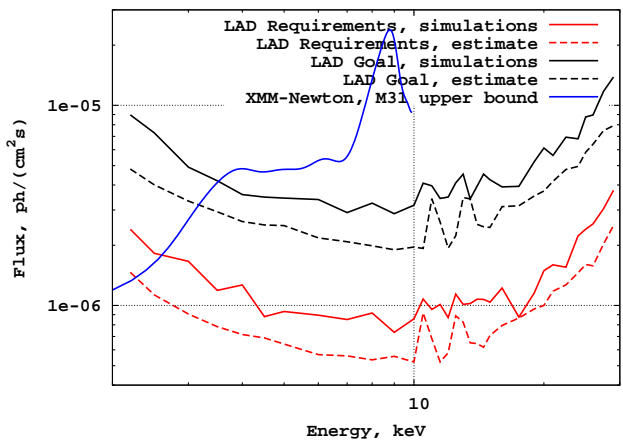}}
\caption{The $3\sigma$ upper bound on the flux in the line from a diffuse
  source detectable by LAD detector. Thick lines: results based on simulations
  and subsequent detection of a line. Dashed line: $3\sigma$ estimates, based
  on the statistical $3\sigma$ upper bound of the instrumental background, see
  Eq.~(\protect\ref{eq:1}). 3$\sigma$ upper bound on DM flux from \xmm\
  observations of M31 central part~\protect\cite{Boyarsky:07a} (blue) is shown
  for comparison. }
  \label{fig:flux_bound}
\end{figure}

\section{Sensitivity  for the DM line detection}
\label{sec:sens-loft-detect}

\subsection{Signal from extended sources in the field of view of collimator}

We begin with an estimate of the sensitivity of the LAD detector for weak
diffuse lines in the energy range below 30~keV where the collimator limits the FoV to $1^\circ$. To this end we take the background spectrum shown in
Fig.~\ref{fig:LOFT-internal-vs-extragal-10ks}, and compute the number of
background photons in the bin with the size equal to FWHM over the time
$T_{exp}$ chosen to be 100~ksec (a typical timescale of a single
observation). We then estimate the $3\sigma$ upper limits on the line flux in
each narrow energy, based on the statistical error on the background counts:
\begin{equation}
 \label{eq:1}
 F_{line,\ 3\sigma}(E)<\frac{3\sqrt{2\times N_{bkg}(E)}}{A_{\eff}(E)T}
\end{equation}
(an additional $\sqrt 2$ was included, because we assumed that we are
subtracting observations from a background with comparable statistics).

The upper limit calculated in this way is shown in Fig.~\ref{fig:flux_bound}. One could see that this limit is better than that derived from an {\it XMM-Newton} exposure of the same duration. This demonstrates that in spite of somewhat higher background level of the LAD detector (contrary to {\it XMM-Newton} it includes the CXB scatterd by the collimator walls), the upper limit on the line flux within the FoV is better. The obvious reason for this is much larger effective area of the detector. Further improvement of sensitivity of LAD, compared to {\it XMM-Netwon} (not reflected by the figure) is that LAD collects larger DM line signal in a similar exposure. This is due to the larger FoV.

\subsection{Signal from the Milky Way halo visible for a ``bare detector''}

At energies above $20-40$~keV, the collimator of the LAD will be not able to
stop photons falling at large incidence angle, so that LAD increasingly
becomes a ``naked detector'' sampling photons from large, steradian scale
FoV. Such a design is optimal for the search of diffuse emission from the
Milky Way halo \cite{Boyarsky:06f,denHerder:09}. The DM signal is accumulated
in all the pointings of the telescope, no matter where the pointing is
directed. This allows to achieve extremely long exposures in a multi-year
operation of the telescope. It is not possible to estimate what will
  be the effective field of view of the LAD detector at these energies. As an
  estimate we take $\Omega_{\fov, \text{high}} = \unit[1]{sr}$. We remind that
  the sensitivity estimate, $R$, scales as $R\propto\sqrt{\Omega_{\fov,
        \text{high}}/\unit[1]{sr}}$.

In the case of an all-sky source, it is a challenge to distinguish the real
source signal from an instrumental feature, such as the instrumental atomic or
nuclear line, which is also expected to appear in all pointings. However, a
clear observational signature of the real DM decay signal is the excess toward
the Galactic Centre (GC). This signature is readily identifiable and could be used
to discriminate the real signal from the instrumental noise. This approach was
used in~\cite{Boyarsky:07b}. The authors of this reference were able to identify for example the 511 keV
from the positron annihilation in the Galactic Centre region. It was also
demonstrated  that no other (instrumental) line present in the all-sky exposure has  surface brightness
profile (as a function of off GC-angle) expected for DM decay line.  This
allowed the authors of~\cite{Boyarsky:07b} to derive constraints on the DM
line flux in the 20 keV -- 7 MeV energy range using the SPI instrument of
INTEGRAL as a wide-field (steradian FoV) detector.

The same approach could be adopted to the LAD data above $\sim 20-40$~keV
where the instrument works as a wide FoV detector. 
The main difference with the calculations of the previous section is that the
central part of the Milky Way is a bright X-ray source. The emission from this
source is the sum of emission from high mass and low mass X-ray binaries and
cataclysmic variables. Measurement of the collective emission from the Milky
Way sources within a steradian scale FoV by SPI~{\cite{Churazov:10}
provides a reference value for the level of sky background on top
of which the DM line signal from the Milky Way should be detected
\begin{equation}
F_{MW}\simeq 10^{-4}\left(\frac{E}{100\mbox{ keV}}\right)^{-2.5}\frac{\mbox{ph}}{\mbox{cm}^2\,\mbox{s}\,\mbox{keV}}.
\label{eq:mw}
\end{equation}

 The limits calculated for the background
level (\ref{eq:mw}) and a year-long exposure time are shown in Fig.~\ref{fig:diffuse_line}. For comparison,
the same figure shows the upper limit on the line flux within a steradian FoV
of SPI found by \cite{Boyarsky:07b}. One could see that, in accordance with
the expectations, the limits which would be derived from the LAD data are
tighter than those from the SPI.

\begin{figure}[!th]
\includegraphics[width=\linewidth]{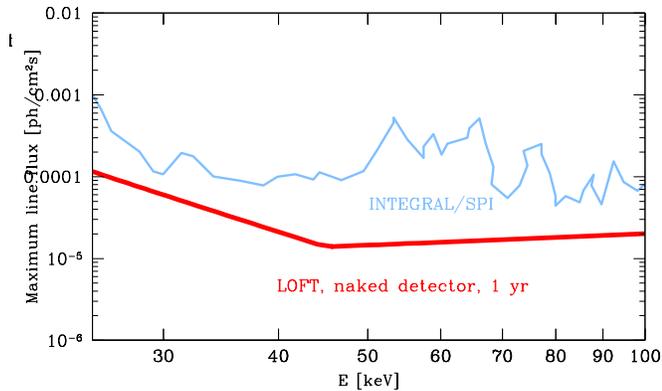}
\caption{Flux limits on DM decay line with large FoV (``bare'') LAD detector
  expected from a year-long observation of diffuse emission within $\Omega \simeq
  1$~sr FoV.  For comparison, the limits found from a long (multi-year)
  exposure of SPI spectrometer on board of INTEGRAL satellite are shown (light
  blue curve). }
\label{fig:diffuse_line}
\end{figure}

\subsection{Limits on the decaying DM lifetime}
\label{sec:limits-decaying-dm}

To convert the limits on the line flux into the limits on the lifetime of
the decaying DM, $\tau_\dm$, we note that flux in line (in photons per $\unit{cm}^2$ per
sec) is given by
\begin{equation}
  \label{eq:flux}
  F_{line} = \parfrac{1}{\tau_\dm m_\dm} \parfrac{M_\fov}{4\pi D_L^2}
\end{equation}
where the first term is determined by the basic properties of DM particles,
while the second one is the characteristic of the object being observed.

For nearby objects that cover the whole FoV of the instrument one can express
\begin{equation}
  \label{eq:4}
  \frac{M_\fov}{4\pi D_L^2} \simeq \frac{S_\dm \Omega_\fov}{4\pi}
\end{equation}
where $S_\dm$ is the average DM column density in a given
direction. This quantity changes very little among objects of different masses
and sizes~\cite{Boyarsky:06c,Boyarsky:09c,Boyarsky:09b} and its typical values
are $\unit[10^{2\div 2.5}]{M_\odot/pc^2}$.  Using this fact and taking into
account that for 2-body decays the mass of DM is related to the energy of
emitted photon via $E_\gamma =\frac 12 M_\dm c^2$, we convert the upper bound
on the flux limit into the lower limit on decaying DM lifetime:
\begin{eqnarray}
  \label{eq:3}
  \tau_{\dm}& =& \frac{S_\dm \Omega_\fov}{8\pi E_\gamma F_{line}}
  \nonumber\\ &\approx& \unit[3.7\times 10^{29}]{sec}\parfrac{S_\dm}{10^2
    M_\odot/\unit{pc^2}}\parfrac{
    \Omega_\fov}{\unit[1]{deg^2}}\nonumber\\&&\parfrac{\unit[10]{keV}}{E_\gamma}\parfrac{\unit[10^{-6}]{ph/sec/cm^2}}{F_{line}}
\end{eqnarray}

From Fig.~\ref{fig:flux_bound} one sees that the upper limits on the line flux
is expected to be at the level of $\unit[10^{-6} -10^{-5}]{ph/cm^2
  /sec}$. Substituting these values into (\ref{eq:flux}) one finds the
sensitivity of the LAD detector at the level $\tau_\dm \sim
\unit[10^{29}]{sec}$ -- at least an order of magnitude better than existing
bounds at these energies. This limit is shown in Fig. \ref{fig:decaying-lifetime} as a function of energy. To estimate the sensitivity in the "naked detector" mode, we assume that the FoV of the detector grows as a powerlaw in the 20-40~keV energy range. Detailed simulations are needed to get a more precise estimate of the opening of the FoV with increasing energy.

\begin{figure}
  \includegraphics[width=\linewidth]{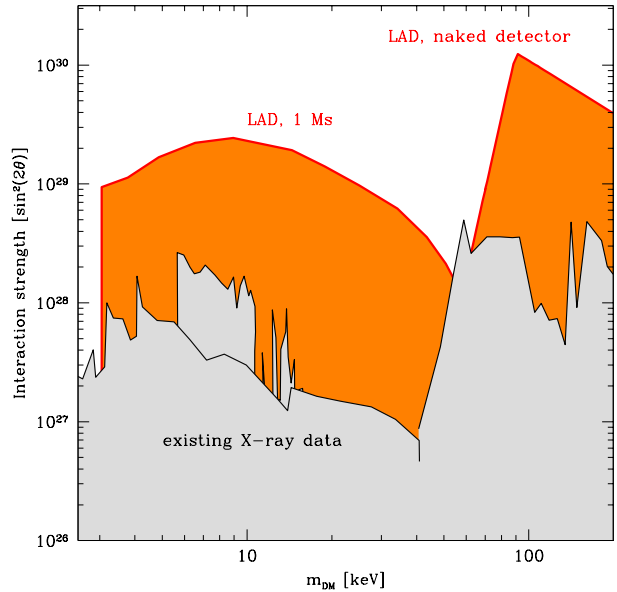}
  \caption{Bounds on lifetime of decaying dark matter (for decays $\text{DM}
    \to \gamma + \nu$ or $\text{DM}\to \gamma + \gamma$) (grey shading) and expected
    improvement from the LOFT LAD detector. Red solid line shows possible LOFT bound
     assuming 1 Ms exposure with the average dark
    matter column density  $S = \unit[300]{M_\odot/pc^{2}}$.}
  \label{fig:decaying-lifetime}
\end{figure}

\section{Implications for sterile neutrino DM models}
\label{sec:implications}

Sterile neutrino  is  a
decaying DM candidate that had recently attracted a lot of attention~(see
e.g.\ \cite{Boyarsky:09a,Kusenko:09a,Boyarsky:12c,Drewes:2013gca} for review).
Sterile neutrino is a right-chiral counterpart of the ordinary (left-chiral)
neutrinos $\nu_e,\nu_\mu,\nu_\tau$. Adding these particles to the SM
Lagrangian makes neutrinos massive and provides a simple and elegant
explanation of the observed neutrino flavor oscillations and of the smallness
of neutrino masses (the so-called ``type I see-saw
model'')~\cite{Minkowski:77,Ramond:79,Mohapatra:79,Yanagida:80}.  These
particles are neutral with respect to all Standard Model interactions (weak,
strong and electromagnetic) (see e.g.\ \cite{Abazajian:2012ys,Boyarsky:12c}
for details). They interact with the matter only
via 
mixing with ordinary neutrinos and in this way effectively participate
in weak reactions~\cite{Boyarsky:12c} with strongly suppressed rate (as
compared to the ordinary neutrinos).  Production of such particles in the
primordial plasma~\cite{Dodelson:93,Shi:98,Abazajian:01a,Asaka:06b,Asaka:06c}
and their decays are controlled by the same parameter -- sterile neutrino
mixing angle $\sin^2(2\theta) \lll 1$ inversely proportional to their
lifetime:
\begin{eqnarray}
  \tau_\dm &=& \frac{1024\pi^{4}}{9 \alpha G_{F}^{2}\sin^{2}( 2\theta)
    m_{\dm}^{5}} \nonumber\\ &\approx& 7.2 \times 10^{29}
  \sec \left[\frac{10^{-8}}{\sin^{2}(2\theta)}\right]\left[\frac{1\keV} {m_\dm}\right]^{5}.
\end{eqnarray}
To be a DM candidate, the interaction strength of sterile neutrinos should be
too feeble to make any sizable contribution to active neutrino
masses~\cite{Boyarsky:06a}. 

The $\nu$MSM model provides an explanation to three known "beyond Standard Model" of particle physics phenomena:  dark matter, baryon asymmetry of the Universe  and
neutrino masses, adding three sterile neutrinos to the Standard
Model particle content~\cite{Asaka:05a,Asaka:05b}.  The lightest of the three sterile neutrinos served as the DM. The combination of X-ray bounds, of primordial abundance results in both upper and lower 
bounds on the mass and mixing angle of DM sterile neutrino in the \numsm. The range of allowed masses of sterile neutrino DM is
$1-50\kev$~\cite{Boyarsky:07b,Boyarsky:09a,Boyarsky:12c}.

The estimates of the bound on the DM sterile neutrino mixing angle expected from LOFT observations are shown in
Fig.~\ref{fig:bounds_combined_dataset_new_LOFT}.  Interestingly,  the``Requirements'' configuration of LOFT is expected to provide the best constraints.
This is mostly due to the fact that the ``Goal'' configuration is optimized
for point sources and therefore LAD FoV is reduced from $\sim 1^\circ$ to
$\sim 0.5^\circ$. This reduces 4 times the expected signal from DM decays
(provided the DM column density is constant across the FoV) while the
background level reduced only slightly.  

One could see that LOFT will be able to explore significant
fraction of the available range of the mixing angles $\theta$ within
$\nu$MSM. Already one  1~Ms long exposure of a dSph galaxy like Ursa Minor
will improve the existing bounds on $\theta$ by two orders of
magnitude. Moreover, taking into account importance of the DM nature problem,
and the unique characteristics of LOFT, which make it an excellent DM
detector, one could imaging a scenario in which the LAD instrument might
  be operated as a dedicated DM detector (e.g. toward the end of the mission),
  accumulating a total year scale exposure of a nearby DM halo. This would
allow a further boost of sensitivity of the detector by a an order of magnitude. In this case LOFT will provide an
almost full test of the $\nu$MSM and either discover the sterile neutrinos or
possibly leave only a narrow window of mass $1$~keV$<m_{wdm}<4$~keV, where the
Ly-$\alpha$ bound suffers from some uncertainties \cite{Boyarsky:09a}, unexplored. To probe the mass range below 4~keV, one might use the LAD data in the energy range below 2~keV. It is clear that the quality of the data in this range is significantly degraded. However, taking into account the unique possibility to explore the full allowed parameter space of a viable DM  model (to find the DM or rule out the model) might serve as a good motivation for the challenging task of data analysis in this energy range. 

\begin{figure}
\includegraphics[width=\linewidth]{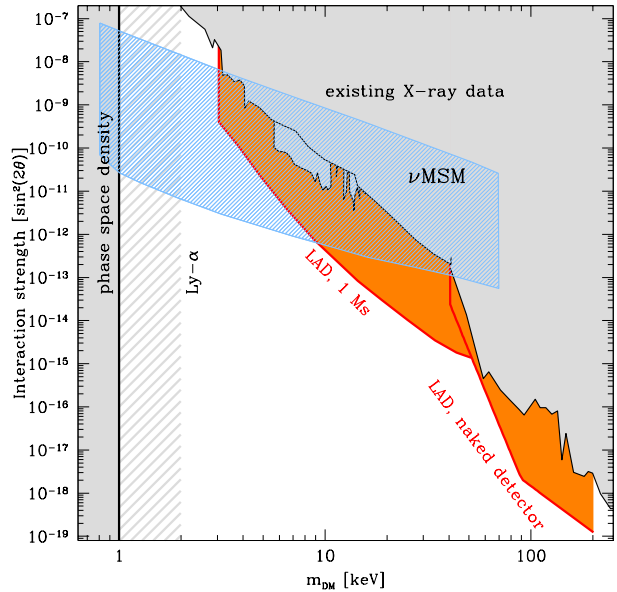}
\caption{Grey shading: Bounds on sterile neutrino parameters. Blue hatching shows the allowed parameter space of  $\nu$MSM model. Orange shading shows the sensitivity limit of LOFT for 1 Ms exposure. }
\label{fig:bounds_combined_dataset_new_LOFT}
\end{figure}

\section{Conclusions}

We have shown that LOFT will be a powerful detector of light decaying DM. 
From Figs. \ref{fig:decaying-lifetime}, \ref{fig:bounds_combined_dataset_new_LOFT} it is clear that LOFT will be one-two orders of magnitude imore sensitive for the detection of DM line in the DM mass range 4-200~keV than all ongoing and past missions. This
will provide a qualitatively new insight into the nature of the DM
particles within various $\Lambda$WDM scenarios, including the most popular one
with sterile neutrino DM. Significant improvement is also expected  at the highest energies above 30~keV, where the LAD instrument becomes a ``naked detector'' with the steradian-scale FoV. Such a
configuration proves to be optimal for search of diffuse all-sky signal from DM decaying  in the Milky Way halo (c.f.~\cite{Boyarsky:06f,denHerder:09}).

The energy range of LOFT is crucially important for testing the reference
$\nu$MSM model. This is clear from Fig. \ref{fig:bounds_combined_dataset_new_LOFT}. If operated as a dedicated DM search experiment, LOFT will be able to probe almost all parameter space of $\nu$MSM.

\bibliography{preamble,LOFT-decayingDM,astro,cmf,combined_numsm,LOFT}

\end{document}